\documentclass[epjST]{svjour}
\usepackage{graphics}
\usepackage{color}
\usepackage{amsmath,amssymb}

\begin{document}
\title{A hierarchical heteroclinic network}
\subtitle{Controlling the time evolution along its paths}
\author{Maximilian Voit\thanks{
  \email{m.voit@jacobs-university.de}}
\and Hildegard Meyer-Ortmanns\thanks{
  \email{h.ortmanns@jacobs-university.de}}}
\institute{Physics and Earth Sciences, Jacobs University Bremen, P.O. Box 750561, 28725
Bremen, Germany}

\abstract{
  We consider a heteroclinic network in the framework of winnerless competition of species. It consists of two levels of heteroclinic cycles. On the lower level, the heteroclinic cycle connects three saddles, each representing the survival of a single species; on the higher level, the cycle connects three such heteroclinic cycles, in which nine species are involved. We show how to tune the predation rates in order to generate the long time scales on the higher level from the shorter time scales on the lower level. Moreover, when we tune a single bifurcation parameter, first the motion along the lower and next along the higher-level heteroclinic cycles are replaced by a heteroclinic cycle between 3-species coexistence-fixed points and by a 9-species coexistence-fixed point, respectively. We also observe a similar impact  of additive noise. Beyond its usual role of preventing the slowing-down of heteroclinic trajectories at small noise level, its increasing strength can replace the lower-level heteroclinic cycle by  3-species coexistence fixed-points, connected by an effective limit cycle, and for even stronger noise the trajectories converge to the 9-species coexistence-fixed point. The model has applications to systems in which slow oscillations modulate fast oscillations with sudden transitions between the temporary winners.
}
\maketitle


\section{Introduction}
\label{introduction}
A heteroclinic network is a sequence of trajectories connecting saddles $\sigma_1,...,\sigma_n$ in a topological network. In particular a heteroclinic cycle  is an invariant set consisting of the union of a set of saddles and the corresponding trajectories, which are backward asymptotic to each saddle $\sigma_i$ and forward asymptotic to $\sigma_{i+1}$ \cite{postleth}. Although this particular nonlinear dynamics looks exceptional, it is frequently found in ordinary differential equations under certain constraints like symmetries \cite{krupa} or delay \cite{kori}. Accordingly it is predicted in models of coupled phase oscillators \cite{kori,ashwin}, vector models \cite{kori}, pulse-coupled oscillators \cite{timme} and models of winnerless competition \cite{valentin1}. The applications range from social systems \cite{hauert,szabo} to ecological systems like foodwebs \cite{valentin1,nowak}, fluid mechanics \cite{tu}, chemostats \cite{hsu}, computation \cite{timme} to neuronal networks
\cite{binding,chunking,rabi1,rabi2,koma1,szucs,koma2,koma3}. In relation to neuronal systems, heteroclinic cycles (and, more generally, heteroclinic sequences) in models of winnerless competition provide mechanisms for generating transient dynamics, which can be sensitive to the very input and robust against perturbations at the same time \cite{valentinchaos14}. The transient feature is appreciated  and favored as compared to infinite-time limits, as realized in attractors like simple fixed points or limit cycles. The reason is that cognitive processes in the brain and episodes in ecological or social systems are inherently transient themselves.

More specifically, for neuronal networks heteroclinic dynamics was proposed as a mechanism of binding between different information modalities in the brain \cite{binding}. Here the winnerless competition via inhibitory connections is between active brain modes, representing the temporal processing of different kind of input. The heteroclinic network stands for multi-dimensional binding options, without implementing any hierarchy. An explicit hierarchy in time scales is implemented in the so-called chunking dynamics of the brain \cite{chunking}. Chunking refers to the phenomenon that the brain uses to perform information processing of long sequences by splitting them into shorter information items, called chunks. In \cite{chunking} the generalized Lotka-Volterra equations only serve as an elementary building block for different levels of the chunking hierarchy, describing the competition between informational items to produce stable sequences of metastable states. The very generation of chunks and the control of the performance of tasks such as a temporally ordered sequence are described by additional equations in \cite{chunking}.

In contrast to this description, the hierarchy in time scales, which we consider here, is merely implemented via the choice of rates in the generalized Lotka-Volterra equations (GLV) on the price that the choice of rates has to be tuned towards some intervals and in a specific order of sizes. From the physics point of view our main interest is in a possible modulation of fast oscillations by slow oscillations as it is a commonly observed feature in brain dynamics. In particular we are interested in how a structural hierarchy in the attractor space (heteroclinic cycles of heteroclinic cycles) can induce such a dynamical generation of time scales. On the two levels of hierarchy which we consider in this paper, the elementary items correspond to the dominance of single  species in 1-species saddles. The short time scales refer to fast oscillations between different saddles within one small heteroclinic cycle (SHC) (analogous to one ``chunk"), while the slow time scale is generated by a large heteroclinic cycle (LHC) between three SHCs.  Our GLV-dynamics is less rich than the chunking dynamics as considered in \cite{chunking}, nevertheless it already allows for a tuning of time scales over orders of magnitude.
So the goal  is then to control the path of the species trajectories through a desired heteroclinic network.

From the structural point of view a remarkable property of this system is that the invariant sets that are the vertices of the LHC are themselves heteroclinic cycles.
Such constructions have been considered before in \cite{ashwinfield}
in the context of depth-two heteroclinic networks.
Our work extends the study of \cite{ashwinfield} by
explicitly providing the very construction of such a network, in view of illustrating the option of dynamically generating different time scales and analyzing the role of noise.

The paper is organized as follows. In section \ref{sec2} we present the model with the main focus on the construction of the predation matrix that ensures the desired LHC of SHCs. Section \ref{sec:results} gives the results on the dynamical generation of a hierarchy in time scales (section \ref{secIII1}) and the reduction of hierarchy levels via tuning the death rate (section \ref{secIII2}) or increasing the noise strength (section \ref{secIII3}). Section \ref{secIV} provides the summary and conclusions.

\section{Construction of the predation matrix}\label{sec2}
\subsection{The model}\label{sec:model}
We study a system of generalized Lotka-Volterra equations, given by
\begin{equation}
  \partial_t s_i = \rho s_i - \gamma s_i^2 - \sum_{j \ne i} A_{i,j} s_i s_j
  \quad i \in \{1,...,9\}\;,
  \label{eq:lvs}
\end{equation}
where $s_i$ denotes the concentration of species $i$,
$\rho$ is the reproduction rate,
$\gamma$ the death rate, and
$A_{i,j}$ the rate by which species $j$ preys on species $i$.
The set of $A_{i,j}$ constitutes the predation matrix $A$.
Without loss of generality we fix the time scale by setting the reproduction rate $\rho = 1$.

Inspired by the predation matrix for rock-paper-scissors(RPS)-games, we choose the predation matrix $A$ as a block matrix.
Its diagonal consists of $3 \times 3$ blocks of the form $m_0$,
similar to the RPS-predation matrix for which $c=1$ and $e=0$.
The off-diagonal blocks $m_d$ have $d$ on their diagonal and $r$ for the
remaining elements, $m_f$ is chosen accordingly.
\begin{align}
  A &= \left(
\begin{array}{ccc}
 m_0 & m_d & m_f \\
 m_f & m_0 & m_d \\
 m_d & m_f & m_0 \\
\end{array}
\right)\;,
\quad\text{where}
  \label{eq:pred-matrix}
  \\
  \nonumber
  m_0 = \left(
\begin{array}{ccc}
 0 & c & e \\
 e & 0 & c \\
 c & e & 0 \\
\end{array}
\right)\;,
&\quad
  m_d = \left(
\begin{array}{ccc}
 d & r & r \\
 r & d & r \\
 r & r & d \\
\end{array}
\right)\;,
\quad
  m_f = \left(
\begin{array}{ccc}
 f & r & r \\
 r & f & r \\
 r & r & f \\
\end{array}
\right)\;.
\end{align}
Note the similarity of the block form of $A$ to the form of $m_0$.
In section \ref{sec:construction} we shall see why we choose this block form and why we keep the entries in the block matrices as independent parameters $c,e,d,f$ and $r$. Although this choice may look rather peculiar, as it stands it still allows a large variety of transients and stationary behavior.

\subsection{Formation of a heteroclinic network}\label{secII1}

\begin{figure}
  \centering
\resizebox{0.5\columnwidth}{!}{%
  \includegraphics{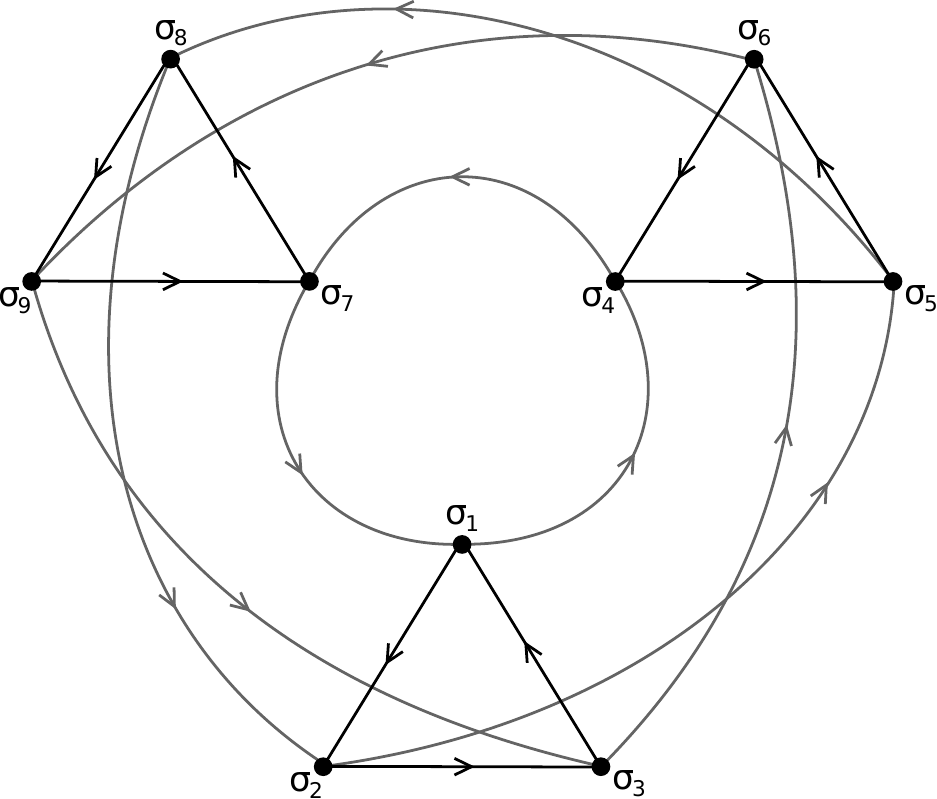} }
\caption{Topology of a hierarchical  heteroclinic network.
  The vertices are marked as black dots.
  For the GLV-system of equation (\ref{eq:lvs}) they correspond to
  single-species fixed points $\sigma_i$, where species $i$ has a population of
  $s_i = \tfrac{\rho}{\gamma}$. Links are heteroclinic connections, arrows mark their direction.
}
  \label{fig:hierhet-topology}
\end{figure}

Our interest is in how much it is possible to control the heteroclinic dynamics by a suitable choice of predation rates $c,e,d,f$ and $r$ so that the vector of nine species evolves between 1-species saddles along a prescribed trajectory in the nine-dimensional phase space, as indicated in Figure \ref{fig:hierhet-topology}. The vertices of the heteroclinic network are 1-species saddles, at which a single species has a non-vanishing concentration, being the temporary ``winner of the game". The links of the network are heteroclinic connections. The rates shall be chosen in such a way that each saddle has two incoming contracting directions and two outgoing repulsive directions of different strength towards which the trajectory can escape. These directions are coming from and going to two other saddles, respectively, while the remaining five directions in phase space should only stabilize the desired trajectories, while moving along one of the indicated  heteroclinic connections.

As indicated in Figure \ref{fig:hierhet-topology}, we search for small heteroclinic cycles (SHCs) between species $1,2,3$, as well as $4,5,6$ and $7,8,9$. The species within the SHCs play then RPS, for which species $1$ preys on $3$, $3$ on $2$, and $2$ on $1$ etc.. (Note that the arrows, indicated along the heteroclinic cycles and following the time evolution of a desired trajectory, are reverse to the predation relations of who preys on whom.)

In addition, we want to have heteroclinic connections also between saddles belonging to different clusters, for example, from saddle $\sigma_2$ to $\sigma_5$, and $\sigma_5$ to $\sigma_ 9$ and back to $\sigma_2$, making up a large heteroclinic cycle (LHC). As indicated in Figure \ref{fig:hierhet-topology}, the directed connections between different SHCs are allowed from each saddle via a single heteroclinic connection. This means that an LHC amounts to a cycle between SHCs, where the connections between the different SHCs may be realized via any of the saddles from an SHC.
The distinction between small and large heteroclinic cycles should not be understood as referring to the distances along heteroclinic connections in phase space, which are all of the same order of magnitude due to the choice of saddle coordinates. In Figure \ref{fig:hierhet-topology} they appear different only due to the projections on two-dimensional space. Instead, it refers to short and long time scales, as we shall choose the rates in a way that typically the trajectory spends a long time within one of the SHCs, before it escapes to the next SHC and altogether repeatedly orbits along LHCs. If we define the long time scale as one complete revolution of an LHC (``complete" in the sense that each of the three SHCs has been visited once), this time is then roughly three times  the number of revolutions within one SHC, as the transition between two SHCs happens almost instantaneously.

Due to the characteristic slowing-down effect of heteroclinic cycles (as long as no noise is applied), it is not meaningful to talk about a characteristic period of an SHC or an LHC. Yet it makes sense to compare the speed with which the winners within a single cluster alternate, with the speed with which whole clusters alternate in dominating the species of the other clusters if both are compared in the same fixed time interval. In this time interval one will then observe fast oscillating species concentrations, modulated by slowly oscillating cluster dominance.  This combination of slow and fast oscillations is shared with chunking dynamics, here, however, to be generated merely by the GLV-equations via a heteroclinic cycle of heteroclinic cycles.

Note that while the system considered in \cite{ashwinfield} is similar in the overall structure  to the present one, there each saddle connects to all other saddles of the successive SHC rather than to one as in our case. The examples discussed in \cite[sections 5 and 6]{ashwinfield} are related to our system as they also possess a $\mathbb{Z}_3 \times \mathbb{Z}_3$ symmetry and a
similar topology. The $\mathbb{Z}_2$ symmetry of the cubic equations in
\cite{ashwinfield} is absent in our system, as are certain degeneracy breaking
terms.

In the following we shall see how we can control an actual trajectory along the hypothetical network of Figure \ref{fig:hierhet-topology}.

\subsection{Fixed point classes}\label{sec:fixedpointclasses}
For the calculation of fixed points, note that either $s_i$ itself must be zero
in equation~(\ref{eq:lvs}), or the bracket term that remains when $s_i$ is factored out.
Thus, there are maximally $2^9 = 512$ fixed points in total,
which can be partitioned into classes according to their permutation symmetry. In Table \ref{tab:fixedpoints} we list the classes which are relevant for the further considerations, together with their general form, coordinates and eigenvalues of the Jacobian, evaluated at these fixed points. $FP_0$ stands for the fixed point, corresponding to global extinction of species. It is characterized by $s_i=0$ for $i\in\{1,...,9\}$, the nine eigenvalues are equal to $\rho>0$, so that it is unstable.

 $FP_1$ stands for single-species fixed points. Their eigenvalues as indicated in the table are in the main focus of our interest. They will be appropriately chosen to guarantee the desired stability properties.

 There are multiple classes of fixed points involving three species. Most relevant is the class containing only those species whose single-species fixed points belong to the same three-cycle, i.e.
$s_i = s_{i+1} = s_{i+2}, s_j = 0 \,\forall j \notin \{i,i+1,i+2\}$ for $i \in \{1,4,7\}$.
These species form what we will call a ``cluster". Accordingly, we denote the class of the (local) three cluster-coexistence fixed points $FP_c$.
Each of them has one pair of complex conjugated eigenvalues, corresponding to two unstable directions for parameters which satisfy equations (\ref{eq:cond-parameters}) below. The remaining eigenvalues are
$-\rho$, $\tfrac{\rho(c+\gamma -d+e-2 r)}{c+\gamma+e}$, and $\frac{\rho  (c+\gamma +e-f-2 r)}{c+\gamma +e}$.
The latter ones, both three times degenerate, are stable for $c+\gamma +e>d+2 r$
and $c+\gamma +e>f+2 r$, respectively.

Finally, there is the class $FP_g$, containing only the global
coexistence fixed point, $s_i = s_1 \,\forall i$.
Here, all species have the same concentration. It has four pairs of complex conjugated eigenvalues.
For parameters according to equation (\ref{eq:cond-parameters}) below, at least one of them has a
positive real part, making this fixed point unstable.
The remaining eigenvalue is $-\rho$.

While both of the latter fixed point classes are not directly involved in the
heteroclinic network, their stability properties affect the dynamics.
All other fixed points (of which one up to seven coordinates are zero) are either not physical or irrelevant in the sense that the desired trajectory does not ``care" about them due to the choice of predation rates that we derive in the next section.

\begin{table}[htb]
  \centering
  \scriptsize
  \caption{Selected relevant fixed points of the nine-species generalized Lotka-Volterra system.}
  \label{tab:fixedpoints}
  \bgroup
  \def\arraystretch{2.3}
  \begin{tabular}{p{1.6cm}|p{4.5cm}|p{5.4cm}}
    class \& name & form / location & eigenvalues
    \\
    \hline
    \hline
    FP0: \textit{extinction} &
    $s_i = 0 \, \forall i$ &
      $\rho~(9\times)$
    \\
    \hline
    FP1: \textit{single species} &
    $s_i = \frac{\rho}{\gamma}, s_j = 0 \, \forall j \ne i$ &
      $-\rho ,\rho -\frac{c \rho }{\gamma },\rho -\frac{d \rho }{\gamma },\rho
      -\frac{e \rho }{\gamma },\rho -\frac{f \rho }{\gamma },\rho -\frac{\rho
      r}{\gamma }~(4\times)$
      \\
    \hline
    FPc:  \textit{cluster coexistence} &
    $s_i = s_{i+1} = s_{i+2} = \frac{\rho}{c+\gamma +e},
    s_j = 0 \,\forall j
      \notin \{i,i+1,i+2\}$ for $i \in \{1,4,7\}$ &
      $\frac{\rho \left(\pm \sqrt{3} \sqrt{-(c-e)^2}+c-2 \gamma +e\right)}{2
        (c+\gamma +e)}$,\;$-\rho$,
        $\frac{\rho (c+\gamma -d+e-2r)}{c+\gamma +e}$~($3\times$),
        $\frac{\rho (c+\gamma -f+e-2r)}{c+\gamma +e}$~($3\times$)
    \\
    \hline
    FPg:  \textit{global coexistence} &
      $s_i = \frac{\rho}{c+\gamma +d+e+f+4r} \,\forall i$ &
      $\frac{\rho  \left(\pm\sqrt{3} \sqrt{-(c+d-e-f)^2}+c+d+e+f-2 (\gamma
        +r)\right)}{2 (c+\gamma +d+e+f+4 r)}$~($4\times$), $-\rho$
  \end{tabular}
  \egroup
\end{table}

\subsection{Choice of the predation rates}
\label{sec:construction}
The choice of the predation rates will be determined by the eigenvalues of the Jacobian, evaluated at the single-species fixed points.
Let us start with the Jacobian, evaluated at the saddle $\sigma_1$ as prototype for 1-species saddles. It is given by the matrix

\begin{align*}
  \left(
\begin{array}{ccccccccc}
 -\rho  & -\frac{e \rho }{\gamma } & -\frac{c \rho }{\gamma } & -\frac{f \rho }{\gamma } & -\frac{r \rho }{\gamma } & -\frac{r \rho }{\gamma } & -\frac{d \rho }{\gamma } & -\frac{r \rho }{\gamma } & -\frac{r \rho }{\gamma } \\
 0 & \rho -\frac{c \rho }{\gamma } & 0 & 0 & 0 & 0 & 0 & 0 & 0 \\
 0 & 0 & \rho -\frac{e \rho }{\gamma } & 0 & 0 & 0 & 0 & 0 & 0 \\
 0 & 0 & 0 & \rho -\frac{d \rho }{\gamma } & 0 & 0 & 0 & 0 & 0 \\
 0 & 0 & 0 & 0 & \rho -\frac{r \rho }{\gamma } & 0 & 0 & 0 & 0 \\
 0 & 0 & 0 & 0 & 0 & \rho -\frac{r \rho }{\gamma } & 0 & 0 & 0 \\
 0 & 0 & 0 & 0 & 0 & 0 & \rho -\frac{f \rho }{\gamma } & 0 & 0 \\
 0 & 0 & 0 & 0 & 0 & 0 & 0 & \rho -\frac{r \rho }{\gamma } & 0 \\
 0 & 0 & 0 & 0 & 0 & 0 & 0 & 0 & \rho -\frac{r \rho }{\gamma } \\
\end{array}
\right).
\end{align*}

Its eigenvalues are listed in Table 1. It is, however, instructive to first write them in terms of general matrix elements $A_{i,j}$ in order to establish their desired properties, which can be read off from the heteroclinic network of Figure 1.  From the perspective of vertex $\sigma_1$, the direction of the eigenvector to vertex $\sigma_2$ should correspond to an expanding direction within the small cycle SHC (es), towards $\sigma_3$ a contracting direction  within the small cycle (cs), towards $\sigma_4$ an expanding direction  within the large cycle (el) and a contracting direction towards $\sigma_7$ within the LHC (cl). We call the vector along the 1-axis the ``radial"-direction, while the remaining four eigenvector directions ``transverse".

Before we formulate the conditions on the eigenvalues to achieve the desired motion, we introduce a notation that is independent of the actual index values. Corresponding to the former distinction between expanding and contracting directions within the small and large heteroclinic cycles and the remaining ones, we define functions $es$, $el$, $cs$, $cl$ and $r$ between indices, such that $es(i)=j$ ($cs(i)=j$) with $j$ the index of the node, towards which the trajectory from $\sigma_i$ expands (from which the motion contracts towards $\sigma_i$), respectively, both within the small cycle. For example, $es(1)=2$ and $cs(1)=3$. Note that $es^3(i)=i$ and $es^2(i)=cs(i)$ for  a 3-cycle. Similarly, $el(i)$ and $cl(i)$ yield the indices of the outgoing and incoming connections in the large cycle, while $r(i)$ denote the indices assigned to $i$ in the four transverse directions ($r(i) := \{1,\dots,9\} \setminus \{i, es(i),el(i),cs(i),cl(i)\}$).

Let us start with the expanding directions, that is, the outgoing heteroclinics from $\sigma_i$ to $\sigma_{es(i)}$ and $\sigma_{el(i)}$. Both directions must be unstable, their eigenvalues positive. In addition we require that the small cycle is preferred over the large one in the sense that the trajectory spends some time within the SHC, before it escapes to the LHC. The inequality between the corresponding eigenvalues translates into
\begin{equation}
  0 < A_{es(i),i} < A_{ec(i),i} < \gamma
  \label{eq:cond-exp}
\end{equation}
as the first condition. The coefficient of $\vec e_i$ in the corresponding eigenvector $\frac{-A_{i,j}}{2\gamma-A_{j,i}}\vec e_i+\vec e_j$ for $j\in\{es(i),el(i)\}$ is then negative, so that the species  concentration $s_i$ decreases when following these directions.

Next we turn to the contracting directions, associated with the incoming heteroclinics that reach $\sigma_i$ from $\sigma_{cs(i)}$ and $\sigma_{cl(i)}$. As they are stable, the corresponding eigenvalues must be negative. In addition, we impose a condition that is based on a conjecture made in \cite{ashwinpost}.
Accordingly, the absolute values of eigenvalues of contracting directions should be larger than that of the strongest expanding direction to achieve asymptotic stability in the heteroclinic network. This means
\begin{equation}
  \left| \rho \frac{\gamma - A_{cs(i),i}}{\gamma} \right| >
  \left| \rho \frac{\gamma - A_{es(i),i}}{\gamma} \right|
  \quad \wedge \quad
  \left| \rho \frac{\gamma - A_{cl(i),i}}{\gamma} \right| >
  \left| \rho \frac{\gamma - A_{es(i),i}}{\gamma} \right|\;,
  \label{eq:cond-contr-helper}
\end{equation}
which results in the conditions
\begin{equation}
  A_{cs(i),i} > 2 \gamma - A_{es(i),i}
  \quad \wedge \quad
  A_{cl(i),i} > 2 \gamma - A_{es(i),i}\;.
  \label{eq:cond-contr}
\end{equation}

Concerning the transverse directions, in \cite{ashwinpost} it is conjectured that a sufficient (but not necessary) condition for asymptotic stability of the embedding heteroclinic network is that in particular all transverse eigenvalues are negative. In our system this leads to
\begin{equation}
  A_{j,i} > \gamma
  \quad \forall j \in r(i)\;.
  \label{eq:cond-transv}
\end{equation}

At this point it is then suggested to introduce as many different parameters in the predation matrix as are needed for satisfying conditions (\ref{eq:cond-exp}), (\ref{eq:cond-contr}), and (\ref{eq:cond-transv}). We define the parameters $e = A_{es(i),i}$,
$f = A_{el(i),i}$,
$c = A_{cs(i),i}$,
$d = A_{cl(i),i}$, and
$r = A_{j,i}$
for all $i$, with $j \in r(i)$ . Thereby we guarantee permutation symmetry
both within the small cycles and the large cycles, i.e., between the three-cycles among each other.
This finally yields the predation matrix equation (\ref{eq:pred-matrix}).
In terms of these parameters, conditions (\ref{eq:cond-exp}), (\ref{eq:cond-contr}), and (\ref{eq:cond-transv})
take the form
\begin{equation}
  0 < e < f < \gamma
  \quad \wedge \quad
  c > 2 \gamma - e
  \quad \wedge \quad
  d > 2 \gamma - e
  \quad \wedge \quad
  r > \gamma\; .
  \label{eq:cond-parameters}
\end{equation}
Note that this uniform choice of parameters (both $c,d,e,f$, and $r$ within the
predation matrix and $\rho$ and $\gamma$ in (\ref{eq:lvs}))
introduces a $\mathbb{Z}_3 \times \mathbb{Z}_3$ symmetry. This may look rather artificial in view of any applications, where such a finetuning of parameters is highly unlikely.
We verified that the original dynamics near a heteroclinic cycle of heteroclinic
cycles persists when breaking this symmetry, see the next section.

Table 2 summarizes the eigenvalues of the Jacobian at the single-species fixed point $\sigma_1$. The function $\nu(x)$ is defined as $\nu(x) = \rho - \tfrac{\rho x}{\gamma}$.
    Numerical values are rounded to two decimals and evaluated  for our standard choice of parameters:
    $\rho = 1, \gamma = 1.07, r = 1.25, e = 0.2, f = 0.3, c = d = 2$.
    As abbreviations for the directions we use ``rad"= radial, ``exp." = expanding, ``ctr." = contracting,
    "l" = large, "s" = small, "trns." = transverse.

\begin{table}[h]
  \centering
  \caption{Table of eigenvalues at the single species fixed point $\sigma_1$. For further explanations see the text.
}
  \label{tab:eigenvalues2}
  \bgroup
  \scriptsize
  \def\arraystretch{1.5}
  \begin{tabular}{l|ccccccccc}
    direction & rad & exp.s & ctr.s & exp.l & trns. & trns. & ctr.l
              & trns. & trns.
    \\
    \hline
    \hline
    general
      & $-\rho $
      & $\nu (A_{2,1})$
      & $\nu (A_{3,1})$
      & $\nu (A_{4,1})$
      & $\nu (A_{5,1})$
      & $\nu (A_{6,1})$
      & $\nu (A_{7,1})$
      & $\nu (A_{8,1})$
      & $\nu (A_{9,1})$
    \\
    \hline
    parameters
      & $-\rho $
      &$\nu(e)$
			&$\nu(c)$
			&$\nu(f)$
			&$\nu(r)$
			&$\nu(r)$
			&$\nu(d)$
			&$\nu(r)$
			&$\nu(r)$
    \\
    \hline
    numeric
			&$-1$
			&$0.81$
			&$-0.87$
			&$0.72$
			&$-0.17$
			&$-0.17$
			&$-0.87$
			&$-0.17$
			&$-0.17$
  \end{tabular}
  \egroup
\end{table}

\section{Results}\label{sec:results}
\subsection{Generating hierarchies in time scales}\label{secIII1}
\begin{figure}[h]
  \centering
\resizebox{0.45\columnwidth}{!}{%
  \includegraphics{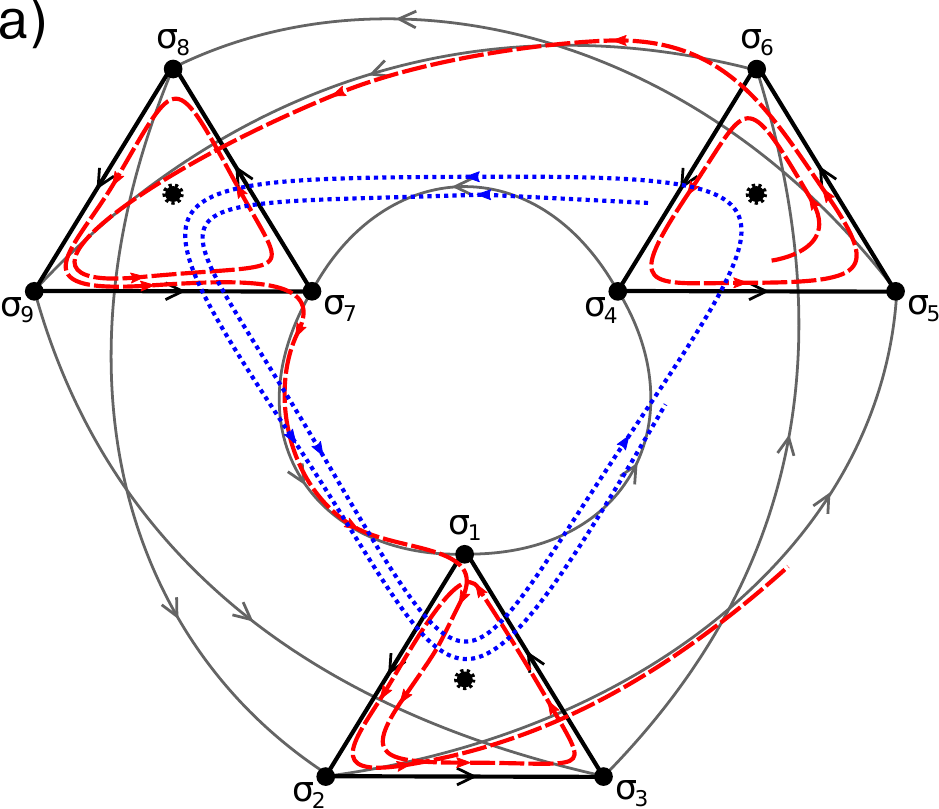}
}
  \hfill
\resizebox{0.45\columnwidth}{!}{%
  \includegraphics{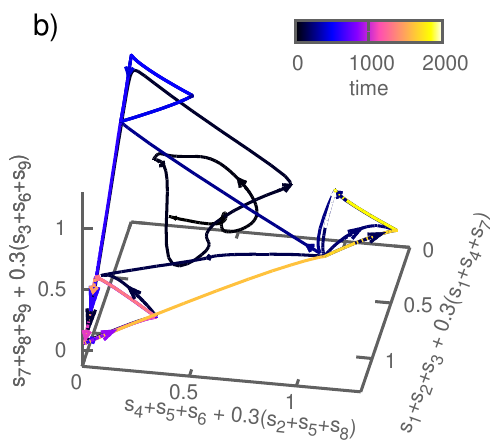}
}
  \caption{Sketch of trajectories in the heteroclinic network with two sample trajectories.
  a) schematic: Solid grey and black lines mark heteroclinic connections between the saddles
  $\sigma_i$. The red curve is a sketch of a trajectory performing LHCs of SHCs on both levels of the hierarchy.
  The blue dashed curve connects cluster-coexistence fixed
  points (dotted), when the lowest hierarchy level is broken as discussed in section \ref{secIII3}.
  b) Plot of a simulated trajectory (cf. fig. \ref{fig:dynamics-2lvl}) in a projection
  of phase space. The color codes time, from black (early) to late (yellow). For further explanations see the text.
}
  \label{fig:phasespace}
\end{figure}

In order to modulate fast oscillations by slow oscillations, we are primarily interested in heteroclinic connections as indicated by the red (long-dashed) trajectory in Figure \ref{fig:phasespace} a), which performs a LHC of SHCs. Starting in the cluster with a SHC with species $4,5,6$, it escapes to the LHC at $\sigma_6$ and later returns to $\sigma_5$, closing the LHC after spending some time within the SHCs. While Figure \ref{fig:phasespace} a) is schematic, Figure \ref{fig:phasespace} b) shows a projection of a simulated LHC of SHCs on a three-dimensional hyperplane of phase space, where the color codes time. Initially the system starts from a randomly chosen location and performs a first LHC (black). At later times, blue, red, yellow colors overwrite the black one in repeated orbits along both cycles. The projection is chosen so as to represent the hierarchy similar to the sketch. Note that the size of the cycles in phase space does not differ as suggested by the Figure. All saddles are equidistant from the origin, so that inter- and intra-cluster distances between saddles are equal in phase space.

\begin{figure}[h]
  \centering
\resizebox{0.99\columnwidth}{!}{%
  \includegraphics{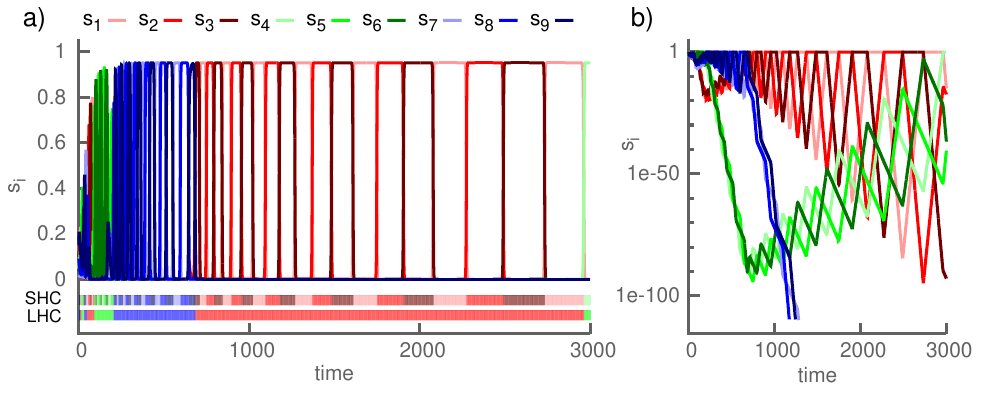}}
\caption{Time evolution according to equation (\ref{eq:lvs}), as a) linear plot
  and b) log-linear plot.
  a) The times during which species $1, 2, 3$ are active reflect the
  periods of activity of that three-cycle.
  Heteroclinic switching occurs both between the three 3-cycles and
  between the species they consist of.  Color bars in a) mark the times
  during which the corresponding species (SHC) or cluster (LHC) is most prominent.
  b) Zoom into the concentrations of non-dominant species during the time period when the red cluster is active. These species keep on competing with each
  other all the time. Parameters are $\rho = 1, \gamma = 1.05, r = 1.25, e = 0.2, f=0.3,$
  and $c = d = 2$.  Initial conditions are sampled uniformly randomly from $[0,0.1]$.
}
  \label{fig:dynamics-2lvl}
\end{figure}

Figure \ref{fig:dynamics-2lvl} shows the actual time evolution of the nine species according to equation (\ref{eq:lvs}) with oscillating concentrations. Panel a) displays the typical time characteristics  of heteroclinic cycles, both for the LHC and the SHCs. Clusters performing the LHC are colored in blue red, green, and species, performing SHCs within one cluster, in different shades of the corresponding cluster color. We see an increase of time intervals both between different shades of one color and between colors. The first one reflects the slowing-down within an SHC, the latter one within an LHC. So, as time goes on, the system spends more and more time in the vicinity of an SHC, for which the species of a certain cluster are dominant, and within that SHC, more and more time close to single-species saddles. So far the simulations are performed without noise. Moreover, it is evident that the switching between saddles and between clusters occurs almost instantaneously. The similar system in \cite[Fig. 9]{ashwinfield} shows comparable dynamics. Panel b) shows the same data with a logarithmic $s_i$ axis to reveal the activities also of species, which do not belong to the dominant cluster (the red one for $t>800$).  Even though these species ($s_4,...,s_9$) are strongly suppressed during the dominance of $s_1,...,s_3$, they go on chasing each other rather than being completely frozen. This is in contrast to chunking dynamics as considered in \cite{chunking}, where the species get frozen.

The same qualitative dynamics as in
Figure~\ref{fig:dynamics-2lvl} is also observed if we break the
$\mathbb{Z}^3 \times \mathbb{Z}^3$ symmetry in the following ways:
We replace the predation matrix
$A \rightarrow \tilde A = A \circ (1 + b W)$
where $\circ$ is the entrywise product,
$b$ the ``strength'' of variation and $W \in \mathbb{R}^{9 \times 9}$ a random matrix (elements chosen from a Gaussian distribution with zero mean and unit variance).
In a similar way we modify the reproduction and death rates for each species
$i$ individually, defining:
$\rho_i = \rho \cdot (1+b u_i)$ and
$\gamma_i = \gamma \cdot (1+b v_i)$, where $u,v \in \mathbb{R}^9$ are random
vectors (chosen from a Gaussian distribution as above).
For both ways of symmetry breaking (either both individually or combined)
the original dynamics near a heteroclinic cycle of heteroclinic cycles
persists as long as the variation $b$ is small enough.
Explorative investigations yield for $b = 0.01$ almost unaffected dynamics,
and noticeable quantitative changes for $b = 0.1$.
Strong variations (e.g. $b = 0.2$) lead to severe changes of the dynamics
when the conditions (\ref{eq:cond-parameters}) are violated and bifurcations
occur.

The time scales of slow oscillations between clusters and short oscillations between single-species saddles can differ by orders of magnitude. In spite of the slowing-down effect without noise, we can compare the time scale say of the first performance of a LHC in Figure \ref{fig:dynamics-2lvl} a) that takes of the order of 3000 time units, while the latest SHC of the red cluster takes some hundred time units.

\begin{figure}[hbt]
  \centering
\resizebox{0.4\columnwidth}{!}{%
  \includegraphics{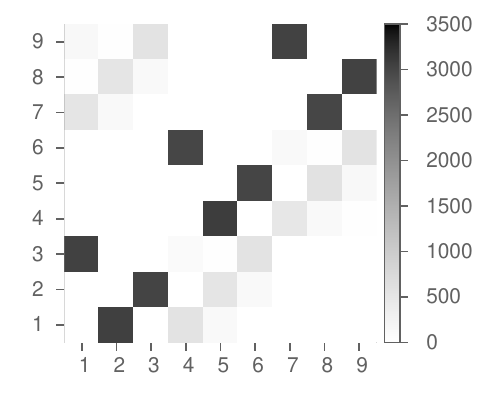}}
\caption{Histogram of transitions between the saddles of the nine species GLV
  system (to be read from row to column).
  Most transitions occur within the small cycles (e.g.  $1 \rightarrow 2$).
  Transitions between three-cycles happen with equal frequency to all
  saddles of the three-cycles (e.g. $1 \rightarrow 4$).  The parameters used are
  $\rho = 1, \gamma = 1.05, r = 1.25, e = 0.2, f=0.3,$ and $c = d = 2$.
  The statistics is collected over 1000 runs of 1000 time steps each.
}
  \label{fig:transition-dhg}
\end{figure}

To further quantify the preference of SHCs over LHCs, we plot  in Figure \ref{fig:transition-dhg} a histogram of transitions between all saddles, for a given fixed choice of parameters. The histogram should be read from columns to rows. So most transitions occur within the small cycles (e.g.  $1\rightarrow 2$ or $6\rightarrow 4$, $9\rightarrow 7$), but not vice versa. We detected saddles $\sigma_i$ as being reached when the corresponding species concentration $s_i$ exceeds a threshold chosen as half of the concentration the fixed point, where it is located. Transitions between three-cycles happen with equal frequency to all saddles of the three-cycles (e.g.  $1 \rightarrow 4$). In addition, a few transitions do not follow the constructed pathway, e.g.
$1 \rightarrow 5$. Here, possibly a transition from 1 to 2 was already in progress, but not detected, so that the actual sequence would have been $1 \rightarrow 2 \rightarrow 5$.

\begin{figure}[htb]
  \centering
\resizebox{0.4\columnwidth}{!}{%
  \includegraphics{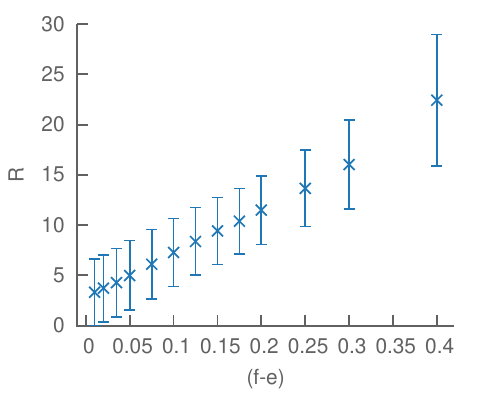}}
\caption{Ratio $R$ of the number of small over large heteroclinic cycles. $R$ is plotted as a function of the difference of parameters $f$ and $e$. Simulations are sampled over 10000 runs of 1000 time steps per data point.
  The other parameters that were used are $\rho = 1, \gamma = 1.05, r = 1.25, e = 0.2,$ and $c = d = 2$.
}
  \label{fig:scf-scaling}
\end{figure}

Next we vary the parameters $e$ and $f$, as it is particularly the choice of these rates that determines the dwell times within an SHC as compared to those at individual saddles and therefore allows a tuning of the preference of SHCs over LHCs. Figure \ref{fig:scf-scaling} shows the ratio $R$ of the number of transitions within small cycles to transitions to other clusters within large cycles, where $f$ was changed, while $e$ was kept fixed.

The reason why we have to collect a statistics in spite of the deterministic dynamics is due to the random choice of initial conditions, uniformly chosen from $[0,0.1]$. Depending on the initial condition, the trajectory enters the immediate neighborhood of one of the saddles, and depending on that location, the number of orbits within an SHC or LHC varies even when all other parameters are kept fixed.

\subsection{Reduction of the hierarchy levels}\label{secIII2}
\begin{figure}[htb]
  \centering
\resizebox{1.0\columnwidth}{!}{%
  \includegraphics{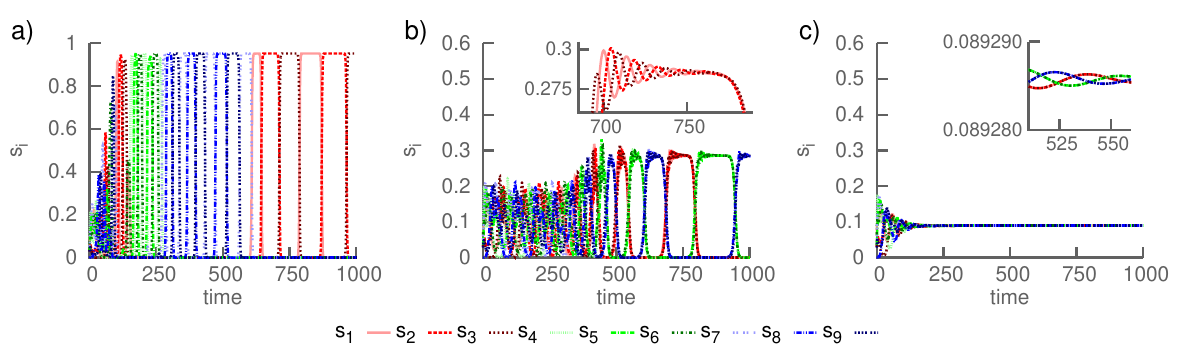}}
\caption{
  Reduction of the hierarchy levels as a function of the death rate $\gamma$.
  Upon increasing $\gamma$, the system goes through a sequence of Hopf bifurcations.
  a) $\gamma = 1.05$ with both hierarchy levels still present; b) $\gamma = 1.3$, where the lowest level of the hierarchy is eliminated and the system is in a heteroclinic cycle between three-species coexistence
  fixed points. c) $\gamma = 1.7$, where the system is at the global coexistence-fixed point.
  The remaining parameters are unchanged:  $\rho = 1, r = 1.25, e = 0.2, f=0.3,$ and $c = d = 2$.
  Initial conditions are sampled uniformly randomly from $[0,0.1]$. For further explanations see the text.
  }
  \label{fig:tuning}
\end{figure}

If we tune the death rate $\gamma$, the system undergoes a sequence of Hopf bifurcations, whose order depends on the concrete choice of parameter values. We distinguish two types of Hopf bifurcations. One of them occurs simultaneously at the different cluster coexistence fixed points at $\gamma^{(c)}$, where the real part in the pair of complex conjugate eigenvalues of the cluster-coexistence fixed points changes sign, so that these fixed points become attracting. As a consequence, the lowest level of the heteroclinic cycle gets extinct, and the dynamics follows a heteroclinic cycle between three 3-species coexistence fixed points $FP_c$, as displayed in Figure \ref{fig:tuning} b) and the blue trajectory in the schematic plot of Figure \ref{fig:phasespace} a).

The other type of Hopf bifurcations happen all at the global coexistence fixed point $FP_g$. Each of them stabilizes eigen-directions, corresponding to one pair of complex conjugated eigenvalues of $FP_g$. In the order of increasing $\gamma$ the last one occurs at
$\gamma_4^{(g)} = \max \, \{
  \tfrac{1}{2} (c+d+e+f-2 r),
  \tfrac{1}{2} (c-2d+e-2f+4r),
  \tfrac{1}{2} (-2 c+d-2 e+f+4 r)
\}$.
As a result of this last bifurcation, the global coexistence fixed point becomes attracting and the system dynamics converges to this fixed point, as shown in Figure \ref{fig:tuning} c). This eliminates the second hierarchy level (slow time scale) from the system.

The order of Hopf bifurcations depends on the concrete parameter choice. For $\rho=1$, $r=1.25$, $e=0.2$ and $c=d=2$, we find
\begin{equation}
0\;<\;\gamma^{(g)}_{1,2}=1.0\; <\; \gamma^{(c)}=1.1\;<\; \gamma_3^{(g)}=1.3\;<\;\gamma_4^{(g)}=1.45\;.
\end{equation}
This means, by tuning one parameter $\gamma$, we are able to reduce the
hierarchy levels of the heteroclinic network from two levels in \ref{fig:tuning}
a) (from $\gamma\ge 1.1$ on) to one in b) and (from $\gamma\ge 1.45$ on) to none
in c). Panels a) and b) correspond to Figures 9 and 10 in \cite{ashwinfield}, respectively.

\subsection{The impact of noise}\label{secIII3}
First we consider additive noise. When the dynamics of a systems follows a heteroclinic cycle, a typical effect of noise is to prevent the slowing-down of the switching events between saddles. Instead, the switching between saddles continues with a period that scales with the logarithm of the noise strength \cite{kori,hansel_clustering_1993}. As a first result on the impact of noise we confirm this scaling behavior on both levels, the switching between species within a cluster and between species belonging to different clusters.

\begin{figure}[htb]
  \centering
\resizebox{0.4\columnwidth}{!}{%
  \includegraphics{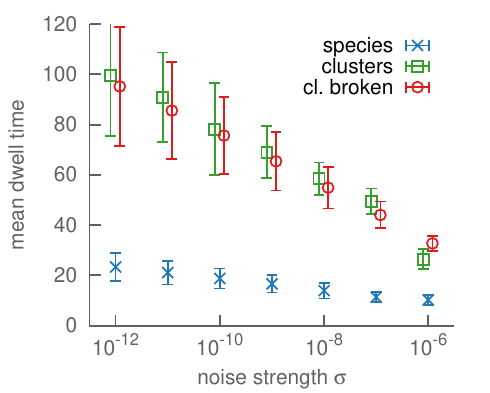}}
\caption{Scaling of the mean dwell-times. Mean dwell-times scale logarithmically
    with the noise strength $\sigma$. Plotted  are the dwell-times of single species (blue) and  of the
    three-cycles (clusters) (green). Dwell times are also shown for the case where the first hierarchy level of SHC is broken and the species coalesce into 3-species coexistence fixed points (red). The dwell-times within an SHC (green) and at a coexistence-fixed point are remarkably similar and drawn at separate noise values only for better readability. Parameters are $\rho = 1, \gamma = 1.05, r = 1.25, e = 0.2, f=0.3,$
    and $c = d = 2$. For the broken hierarchy system $\gamma = 1.3$.
}
  \label{fig:mean-dwelltime}
\end{figure}

We study the influence of noise on equation (\ref{eq:lvs}) by including
additive noise $\xi_i$ with strength $\sigma$.
Since species concentrations must be positive, we take its absolute value,
so the dynamics is given by
\begin{equation}
  \partial_t s_i = \rho s_i - \gamma s_i^2 - \sum_{j \ne i} A_{i,j} s_i s_j
  + \sigma | \xi_i(t) |
  \quad i \in \{1,...,9\},
  \label{eq:lvs-noise}
\end{equation}
where $\xi_i$ itself is Gaussian white noise with zero mean.

The dynamics of equation (\ref{eq:lvs-noise}) leads to periodic switching both on the lowest
hierarchy level and the second level. To further quantify the switching events,
we define as dwell-time the time the system stays in the neighborhood of a
saddle. Here, a viable definition of the neighborhood is the range, where the species concentration
exceeds a threshold $\theta$. For $\theta = 0.4$ (approximately half the full
amplitude of the oscillation), we measure the mean dwell-time of species as
depicted in Figure \ref{fig:mean-dwelltime}.
It is linearly related to the logarithm of the noise strength.

The definition of dwell-time can be extended to three-cycles by
defining the cluster concentrations $S_i = s_{3i-2} + s_{3i-1} + s_{3i}$
for $i \in \{1,2,3\}$.
The system is near cluster $i$ when $S_i > \theta$
(here we use $\theta = 0.5$).
The so-defined dwell-time in the LHC is  also linearly related to the logarithm of the noise strength
(cf. Figure \ref{fig:mean-dwelltime}).
This is in agreement with the fact that the dynamics on the higher level is itself a heteroclinic cycle (of heteroclinic cycles on the lower level).

\begin{figure}[htb]
  \centering
\resizebox{1.0\columnwidth}{!}{%
  \includegraphics{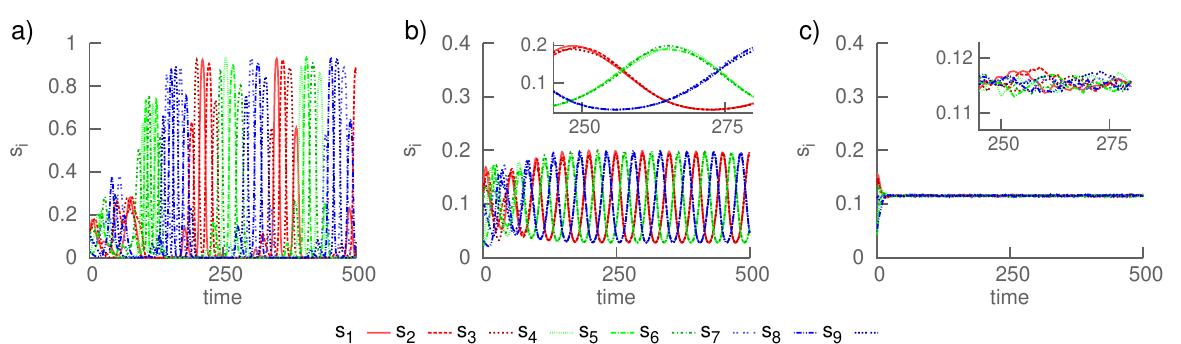}}
\caption{
  Reduction of hierarchy levels by means of an increasing noise strength $\sigma$.
  Upon increasing noise from $\sigma = 10^{-7}$ in a) to $\sigma = 10^{-4}$ in b), the SHC is replaced by three 3-species coexistence fixed points, connected by an effective limit cycle.
  Further increasing the noise strength from $\sigma = 10^{-4}$ in b) to $\sigma = 10^{-3}$ in c)
  drives the system close to the global coexistence-fixed point.
  Parameters are  $\rho = 1, \gamma = 1.05, r = 1.25, e = 0.2, f=0.3,$ and $c = d = 2$.
  Initial conditions are sampled uniformly randomly from $[0,0.1]$.
  }
  \label{fig:tuning-noise}
\end{figure}

In Figure \ref{fig:tuning-noise} we explore the effect of varying the noise strength. First of all note that for Figure \ref{fig:tuning-noise} a) with weak noise, the slowing down of the oscillation frequency is gone.

Beyond this usual effect of noise on heteroclinic motion, in our system it can act similarly to the bifurcation parameter $\gamma$ in the previous section, when we tune its strength to higher values, see Figure \ref{fig:tuning}. For $\sigma\ge 8.5\cdot 10^{-5}$ noise leads to the extinction of the lowest level of heteroclinic cycles; instead the noisy counterparts of the 3-species coexistence-fixed points get connected in an effective limit cycle, see Figure \ref{fig:tuning-noise} b), while for $\sigma\ge 1.5\cdot 10^{-4}$ the effective limit cycle gets eliminated as well, and the dynamics is constrained to the vicinity of the global coexistence-fixed point. The threshold values  of noise, at which the crossover in its effect happens, (that is, from Figure \ref{fig:tuning-noise} a) to b) and b) to c), respectively,) have been determined  by measuring the variance in sums of cluster concentrations in comparison to the sum of variances within single cluster concentrations. First the sum of variances should drastically drop, when the three species  of the three clusters coalesce  to 3-species coexistence-fixed points; next the variance of the sum of different cluster concentrations should drop, when all species coalesce to the global coexistence-fixed point. This is precisely what we observed. It should be remarked that this effect of noise results from an implementation according to equation (\ref{eq:lvs-noise}), that is, with effectively nonzero mean. If we only discard negative  noise contributions from unconstrained Gaussian white noise, once it leads to negative concentrations, an intermediate phase  similar to Figure \ref{fig:tuning-noise} b) is absent.

Next we discuss the role of multiplicative noise as it would be naturally implemented in view of biological applications. If we replace the noise term in equation~(\ref{eq:lvs-noise}) by $\sigma s_i \xi_i(t)$, up to a certain strength its main effect is to smear out the single
species fixed points, which can be understood as follows.
The usual slowing down of the dynamics near a heteroclinic cycle is due to the trajectory
coming closer and closer to the saddles, where the dynamics would get stuck.
Noise increments parallel to the unstable direction of the saddle prevent this
slowing down. However, with multiplicative noise such noise increments are extremely
small, as the concentrations of non-dominant species are small. As a result,
multiplicative noise neither prevents the slowing down nor does it lead to the collapse
of hierarchy levels.

\section{Summary and conclusions}\label{secIV}
Within the generalized Lotka-Volterra model with nine species we have demonstrated how a suitable choice of predation rates can lead to a heteroclinic cycle of heteroclinic cycles, each of which connects three saddles of single species dominance. The different levels of heteroclinic cycles can go along with a dynamical generation of a hierarchy in time scales, as the time it takes a full revolution on the upper level is determined by the dwell-times within the lower-level cycles. In turn, this dwell time depends on the dwell-times in the vicinity of the individual saddles on the lowest level and the number of revolutions within the small heteroclinic cycle. The time scales can mutually differ by an order of magnitude. This feature may be of interest in view of hierarchical time scales observed in the brain, where slow oscillations modulate fast oscillations and external (sensory) input may select a pattern of ``predation" rates in our effective description of transient dynamics such that, for example, chunking is induced. Our choice of rates is specific but not singular as some explicit symmetry breaking in the predation matrix and some perturbation around the choice of reproduction and death rates still support the heteroclinic dynamics.

In view of ecological systems this kind of heteroclinic dynamics would mean that members within a population compete with each other, while simultaneously also clusters of populations compete, resulting in particular in alternating, suddenly changing dominance of a whole cluster over other clusters of populations. It is well known that rock-paper-scissors is played on many scales, starting from the genetic to the cellular and macroscopic level. To our knowledge it is an interesting and open question when in real systems the same game on different scales can be traced back to one and the same underlying microscopic dynamics (as in our artificial system).
In ongoing work we generalize the predation rules towards further levels of hierarchy in heteroclinic cycles and explore the impact of time scales on spatial scales, when we assign the heteroclinic networks to a spatial grid and couple them.

\paragraph{Acknowledgments}
\begin{acknowledgement}
We would like to thank Darka Labavi\'c (University of Lille) for valuable discussions in the beginning of this work.
Also we thank an anonymous referee for drawing our attention to reference \cite{ashwinfield}.
Financial support from the German Research Foundation (DFG, contract ME-1332/28-1) is gratefully acknowledged.
\end{acknowledgement}

\paragraph{Author contributions}
\begin{acknowledgement}
H.M-O. and M.V.  designed the model, discussed the results and contributed to the manuscript. M.V. carried out the numerical experiments and analytical calculations. H.M.-O. proposed the project and wrote the main manuscript text in its final form. Both authors approved the final version.
\end{acknowledgement}

\end{document}